\documentstyle[aasms4]{article}
\newcommand{\beq}{\begin{equation}}
\newcommand{\eeq}{\end{equation}}
\newcommand{\etal}{{\sl et~al.~}}
\newcommand{\kms}{km s$^{-1}$~}
\def\fdg{\hbox{$.\!\!^\circ$}}

\begin{document}

\title{Astrometry with Hubble Space Telescope:
A Parallax of the Fundamental Distance Calibrator $\delta$ Cephei \footnote{Based on 
observations made with
the NASA/ESA Hubble Space Telescope, obtained at the Space Telescope
Science Institute, which is operated by the
Association of Universities for Research in Astronomy, Inc., under NASA
contract NAS5-26555}}

\author{ G.\ Fritz Benedict\altaffilmark{1}, B. E.
McArthur\altaffilmark{1}, L.W.\
Fredrick\altaffilmark{12}, T. E. Harrison\altaffilmark{13}, C. L. Slesnick\altaffilmark{12}, J. Rhee\altaffilmark{12}, \\ R. J. Patterson\altaffilmark{12}, M. F. Skrutskie\altaffilmark{12}
O. G. Franz\altaffilmark{2}, L.\ H. Wasserman\altaffilmark{2}, W.\ H.\ Jefferys\altaffilmark{6}, \\ E.\ Nelan\altaffilmark{5},  W.~van~Altena\altaffilmark{7}, P.~J.~Shelus\altaffilmark{1},
 P.D. Hemenway\altaffilmark{8}, R. L.
Duncombe\altaffilmark{9}, D. Story\altaffilmark{10},  \\ A.\ L.\
Whipple\altaffilmark{10}, and A. J. Bradley\altaffilmark{11} }

\altaffiltext{1}{McDonald Observatory, University of Texas, Austin, TX 78712}
\altaffiltext{2}{Lowell Observatory, 1400 West Mars Hill Rd., Flagstaff, AZ 86001}
\altaffiltext{5}{Space Telescope Science Institute, 3700 San Martin Dr., Baltimore, MD 21218}
\altaffiltext{6}{ Department of Astronomy, University of Texas, Austin, TX 78712}
\altaffiltext{7}{ Department of Astronomy, Yale University, PO Box 208101, New Haven, CT 06520}
 \altaffiltext{8}{Department of Oceanography, University of Rhode Island, Kingston, RI 02881}
\altaffiltext{9}{Department of Aerospace Engineering, University of Texas, Austin, TX 78712}
\altaffiltext{10}{ Jackson and Tull, Aerospace Engineering Division
7375 Executive Place, Suite 200, Seabrook, Md.  20706}
\altaffiltext{11}{Spacecraft System Engineering Services, PO Box 91, Annapolis Junction, MD 20706}
\altaffiltext{12}{ Department of Astronomy, University of Virginia, PO Box 3818, Charlottesville, VA 22903}
\altaffiltext{13}{Department of Astronomy, New Mexico State University, Las Cruces, New Mexico 88003}



\begin{abstract}
We present an absolute parallax and relative proper motion for the fundamental distance scale calibrator, $\delta$ Cep. We obtain these with astrometric data from
FGS 3, a white-light interferometer on {\it HST}. Utilizing spectrophotometric estimates of the absolute parallaxes of our astrometric reference stars and constraining $\delta$ Cep and reference star HD 213307 to belong to the same association (Cep OB6, de Zeeuw et al. 1999), we find $\pi_{abs}  =  3.66 \pm 0.15$ mas. The larger than typical astrometric residuals for the nearby astrometric reference star HD 213307 are found to satisfy Keplerian motion with P = 1.07 $ \pm $ 0.02 years, a perturbation and period that could be due to a F0V companion $\sim7$ mas distant from and $\sim$4 magnitudes fainter than the primary. Spectral classifications and VRIJHKT$_2$M and DDO51 photometry of the astrometric reference frame surrounding $\delta$ Cep indicate that field extinction is high and variable along this line of sight. However the extinction suffered by the reference star nearest (in angular separation and distance) to $\delta$ Cep, HD 213307, is lower and nearly the same as for $\delta$ Cep. Correcting for color differences, we find $<$A$_V>$ = 0.23 $ \pm $ 0.03 for $\delta$ Cep, hence, an absolute magnitude M$_V  =  -3.47  \pm 0.10$. Adopting an average V magnitude, $<$V$>$ =  15.03 $\pm$ 0.03, for Cepheids with log P  =  0.73 in the LMC from Udalski et al. (1999), we find a V-band distance modulus for the LMC, m-M $ =  18.50 \pm 0.13$ or, $18.58 \pm 0.15$, where the latter value results from a highly uncertain metallicity correction (Freedman et al. 2001). These agree with our previous RR Lyr {\it HST} parallax-based determination of the distance modulus of the LMC.

\end{abstract}




%

\section{Introduction}

Many of the methods used to determine the distances to remote galaxies and ultimately the size, age, and shape of the Universe itself depend on our knowledge of the distances to local objects. The most important of these are the Cepheid variable stars. Considerable effort has gone into determining the absolute magnitudes, M$_V$, of these objects (see the comprehensive review by Feast, 1999). Given that the distances of all local Cepheids, except Polaris, are in excess of 250pc, most of these M$_V$ determinations rely on large-number statistics, for example 
\cite{Groen00}, \cite{Lanoix99}, and \cite{Feast97}. Gieren et al. (1993) used Cepheid surface brightness to estimate distances and absolute magnitudes. For Cepheid variables, these determinations are 
complicated by dependence on color and metallicity. Only recently have relatively high-precision trigonometric parallaxes been available for a very few Cepheids (the prototype, $\delta$ Cep and Polaris) from {\it HIPPARCOS} ($\delta$ Cep  =   HIP 110991, \cite{Per97}). We have determined the parallax of $\delta$ Cep with FGS 3 on {\it Hubble Space Telescope} with significantly higher precision. Additionally, our extensive investigation of the astrometric reference stars provides an independent estimation of the line of sight extinction to $\delta$ Cep, a significant contributor to the uncertainty in its absolute magnitude, M$_V$.

In this paper we briefly discuss (Section~\ref{AstRefs}) data acquisition, analysis, and an improved FGS 3 calibration; present the results of spectrophotometry of the astrometric reference stars required to correct our relative parallax to absolute (Section~\ref{SpPhot}); and derive an absolute parallax for $\delta$ Cep (Section~\ref{AbsPi}). Finally, we calculate an absolute magnitude for $\delta$ Cep (Section~\ref{AV1}) and apply it to derive a distance modulus for the Large Magellanic Cloud (Section~\ref{LMCDM}). 

\cite{Bra91} and \cite{Nel01}
provide an overview of the
FGS 3 instrument and \cite{Ben99} describe the fringe tracking (POS) mode astrometric capabilities 
of FGS 3, along with data acquisition and reduction strategies used in the present study. 
We time-tag our data with a modified Julian Date, mJD  =  JD - 2444000.5.

\section{Observations and Data Reduction}  \label{AstRefs}

\begin{figure}
\epsscale{1.0}
\plotone{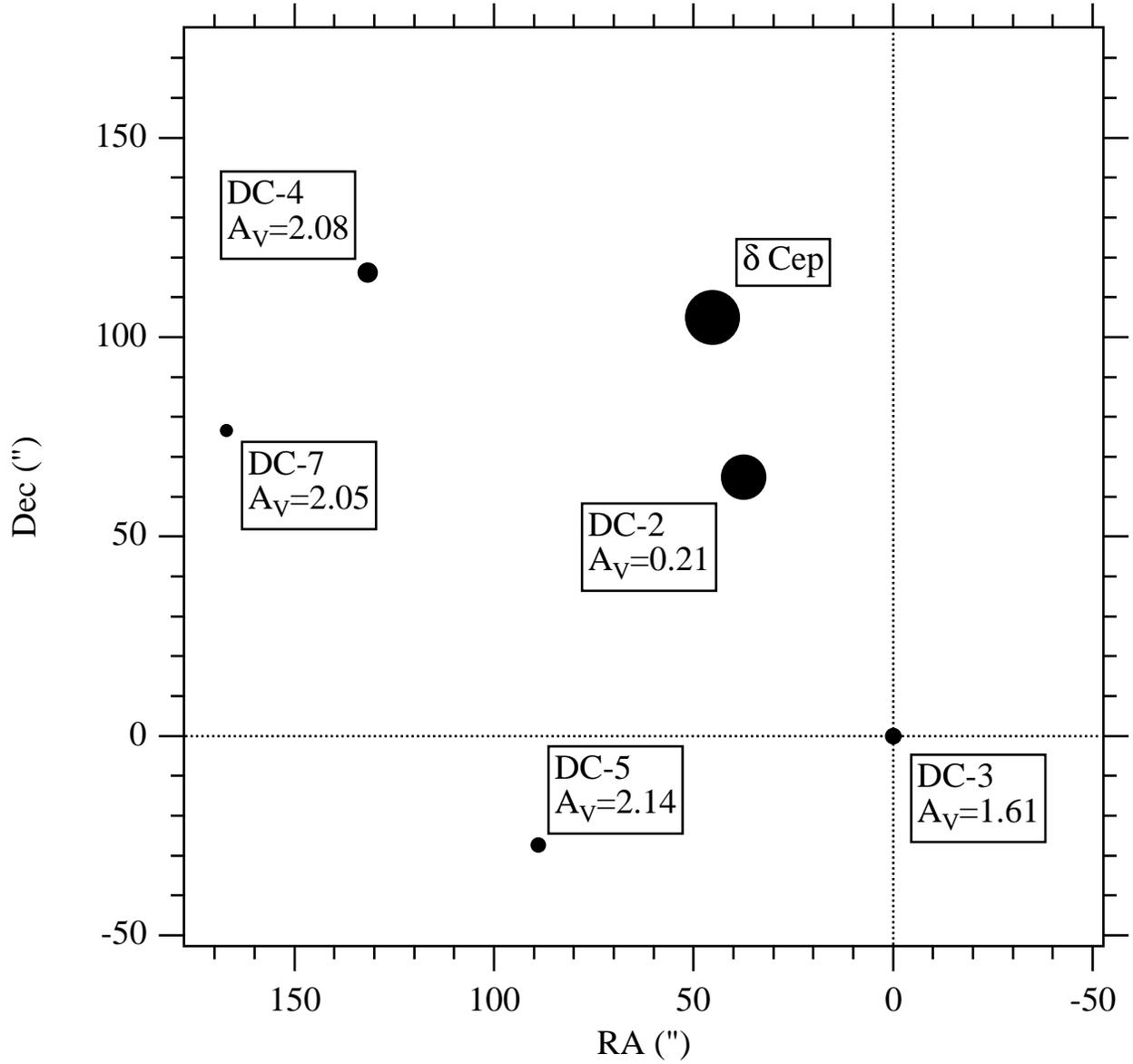}
\caption{$\delta$ Cep and astrometric reference stars. Symbol size is indicative of V magnitude (Table~\ref{tbl-VIS}). The numbers within each identification box are the per-star $<$A$_V>$ from Table \ref{tbl-AV}, Section \ref{AV} and E(B-V) per 100 pc from Table~\ref{tbl-SPP}.}
\label{fig-1}
\end{figure}

Figure \ref{fig-1} shows the distribution in RA and Dec of the five reference stars and $\delta$ Cep. Seven sets of data were acquired, spanning 2.44 years, for a total of 127 measurements  of  $\delta$ Cep and reference stars. Each data set required approximately 40 minutes of spacecraft time. The data were reduced and calibrated as detailed in \cite{Ben99} and \cite{mca01}. In a recent paper (\cite{Ben02}) we described a technique used for these data in which we employ a neutral density filter to relate astrometry of very bright targets to faint reference stars. At each epoch we measured reference stars and the target, $\delta$ Cep, multiple times. We do this to correct for intra-orbit drift of the type seen in the cross filter calibration data in our recent paper reporting a parallax for RR Lyr (\cite{Ben02}, Fig. 1). Data sets 2 and 5 were each also afflicted with an episode of non-monotonic drift, possibly due to mechanism (filter wheel) motion. Fortunately, the strategy of multiple repeats of the observation sequence within each data set permitted generally satisfactory correction. 

All of these observations were acquired with FGS 3, and all before 1998. Since 1999, the prime astrometer aboard {\it HST} has been FGS 1r, installed during the 1999 servicing mission. To calibrate the Optical Field Angle Distortion (OFAD) for FGS 1r we have, during the last three years, secured over 30 observation sets of M35, our calibration target field. These new observations have extended our time base to over ten years and allowed us to refine the proper motions of calibration stars in M35. This has resulted in a more precise star catalog, which in turn has allowed us to improve our FGS 3 OFAD. Applying this revised calibration to the Barnard's Star data presented in \cite{Ben99}, we find a parallax difference of 0.07 mas and a proper motion difference of 0.7 mas yr$^{-1}$, each a change of about one part in 10,000. We use this new calibration for these $\delta$ Cep data.

\setcounter{footnote}{0}
\section{Spectrophotometric Absolute Parallaxes of the Astrometric Reference Stars} \label{SpPhot}
Because the parallax determined for $\delta$ Cep will be
measured with respect to reference frame stars which have their own
parallaxes, we must either apply a statistically derived correction from relative to absolute parallax (van Altena, Lee \& Hoffleit 1995, hereafter YPC95) or, preferably,  estimate the absolute parallaxes of the reference frame stars seen in Figure \ref{fig-1}. With colors, spectral type, and luminosity class for a star one can estimate the absolute magnitude, M$_V$, and V-band absorption, A$_V$. The absolute parallax is then,
\beq
\pi_{abs}  =  10^{-(V-M_V+5-A_V)/5}
\eeq

The luminosity class is generally more difficult to determine than the spectral type (temperature class). However, the derived absolute magnitudes are critically dependent on the luminosity class. To confirm the luminosity classes we employ the technique used by Majewski et al. (2000) to discriminate between giants and dwarfs for stars later than $\sim$ G5, an approach whose theoretical underpinnings are  discussed by \cite{Pal94}.

\subsection{Photometry}
Our band-passes for reference star photometry include: BVRI, JHK (from preliminary 2MASS\footnote{ The Two Micron All Sky Survey
is a joint project of the University of Massachusetts and the Infrared Processing
and Analysis Center/California Institute of Technology } data), and Washington/DDO filters M, 51, and T$_2$ (obtained at McDonald Observatory with the 0.8m Prime Focus Camera).  We transform the 2MASS JHK to the Bessell (1988) system using the transformations provided in \cite{Car01}. In Tables \ref{tbl-VIS} and \ref{tbl-IR} we list the visible, infrared, and Washington/DDO photometry for the $\delta$ Cep  reference stars, DC-2 through DC-7. DC-2 was too bright for 2MASS and our Washington/DDO and RI photometric techniques.

\subsection{Spectroscopy}
The spectra from which we estimated spectral type and luminosity class come from the New Mexico State University Apache 
Point Observatory\footnote{ The
Apache Point Observatory 3.5 m telescope is owned and operated by
the Astrophysical Research Consortium.}. For all but reference star DC-2 classifications were obtained by a combination of template matching and line ratios. Differing classifications for reference star DC-2 (HD 213307) have been reported in the literature. We consider the classifications of both \cite{Lut77} and Savage et al. (1985)\nocite{Sav85}. Table \ref{tbl-SPP} contains a list of the spectral types and luminosity classes for our reference stars, rank ordered by estimated distance. We discuss our estimation of the $<$A$_V>$ in the next subsection.

\begin{figure}
\epsscale{1.0}
\plotone{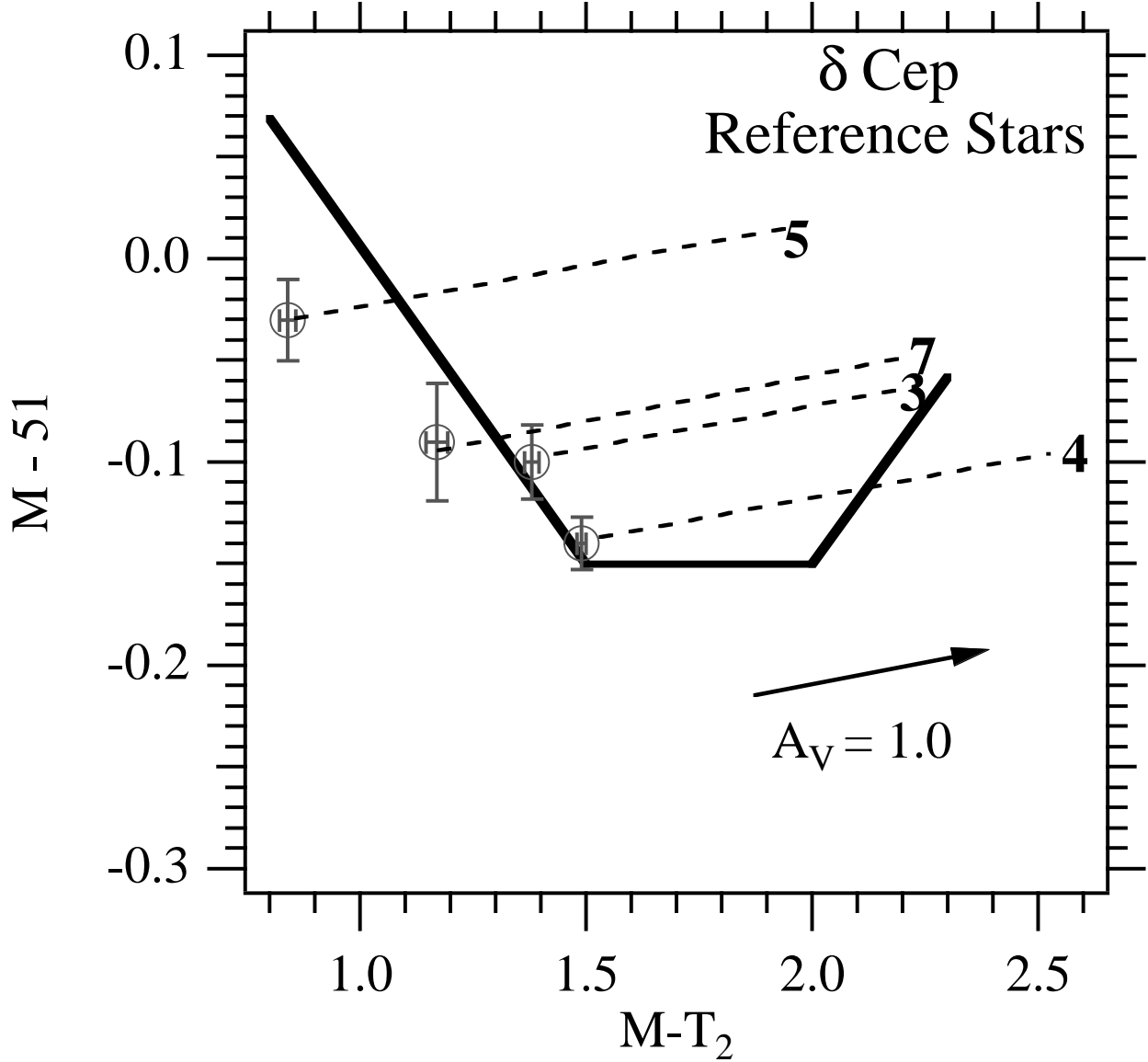}
\caption{M-DDO51 (M-51) vs M-T$_2$ color-color diagram for reference stars DC-3 through DC-7. The solid line is the division between luminosity class V and luminosity class III stars. Giants are above the line, dwarfs below. The reddening vector is for A$_V$ = 1.0. The numbers are the reference star ID's plotted at the observed values. The circles are de-reddened values, based on the per-star $<$A$_V>$ from Table~\ref{tbl-AV}.}
\label{fig-2}
\end{figure}

In Figure \ref{fig-2} we plot the Washington-DDO photometry along with a dividing line between dwarfs and giants (Paltoglou \& Bell 1994 \nocite{Pal94}). The boundary between giants and dwarfs is actually far 'fuzzier' than suggested by the solid line and complicated by the photometric transition from dwarfs to giants through subgiants.  This soft boundary is readily apparent in Majewski et al. (2000) Fig. 14. Objects just above the heavy line are statistically more likely to be giants than objects just below the line. Reference stars DC-2 and DC-5 have spectral types that are too early for this discriminant to work properly. The photometry is consistent with a giant or subgiant classification for the other reference stars.

\subsection{Interstellar Extinction} \label{AV}
\begin{figure}
\epsscale{0.65}
\plotone{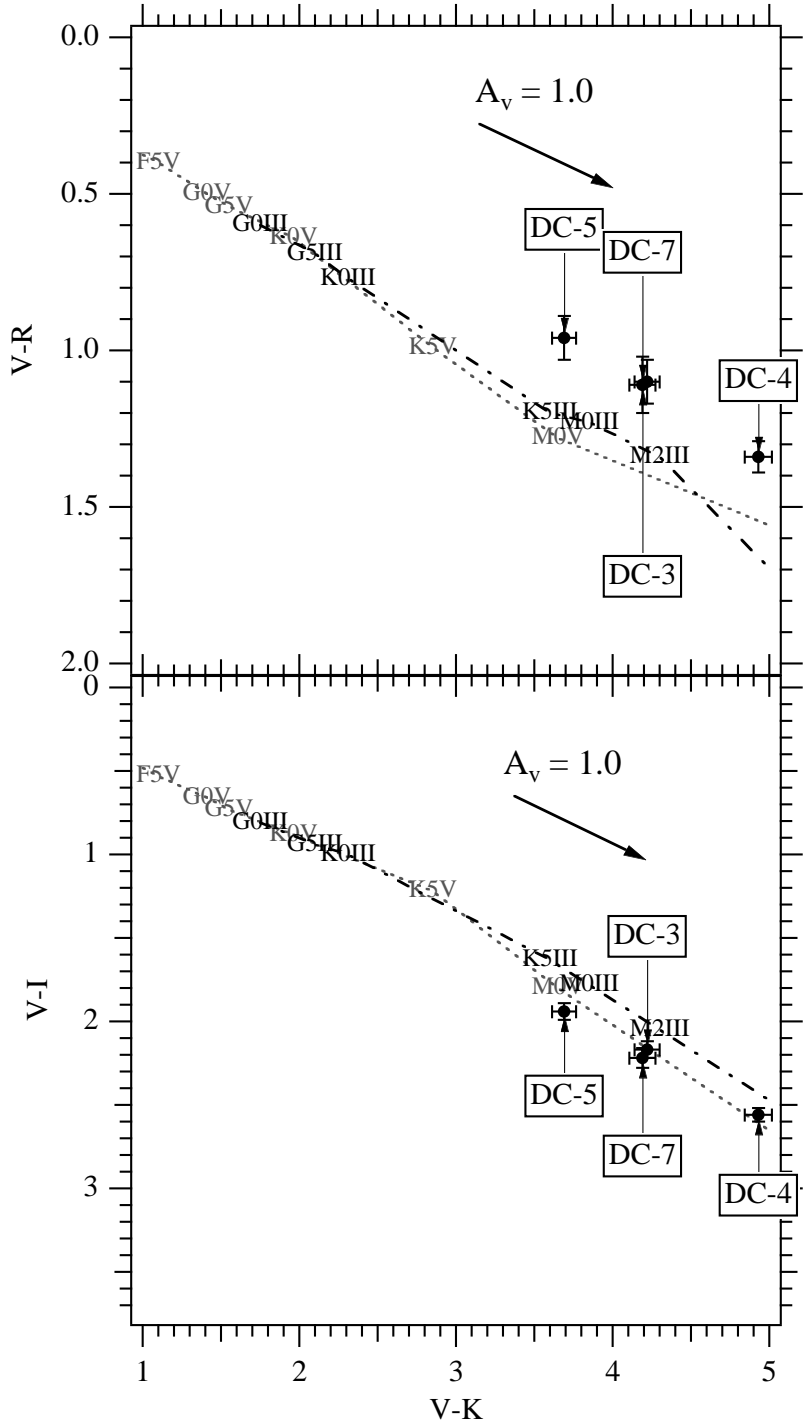}
\caption{V-R vs V-K and V-I vs V-K color-color diagrams for reference stars DC-3 through DC-7. The dashed line is the locus of dwarf (luminosity class V) stars of various spectral types; the dot-dashed line is for giants (luminosity class III). The reddening vector is for an A$_V$ = 1.}
\label{fig-3}
\end{figure}

To determine interstellar extinction we first plot these stars on several color-color diagrams. A comparison of the relationships between spectral type and intrinsic color against measured colors provides an estimate of reddening. Figure \ref{fig-3} contains  V-R vs V-K and V-I vs V-K color-color diagrams and reddening vectors for A$_V$ = 1.0. Also plotted are mappings between spectral type and luminosity class V and III from \cite{Bes88} and \nocite{Cox00} Cox (2000, hereafter AQ00), again with reddening vectors and the loci of luminosity classes V and III stars. Figure~ \ref{fig-3}, along with the estimated spectral types, provides measures of the reddening for each reference star. 

Assuming an R  =  3.1 galactic reddening law (Savage \& Mathis 1977\nocite{Sav79}), we derive A$_V$ values by comparing the measured colors (Tables~\ref{tbl-VIS} and \ref{tbl-IR}) with intrinsic V-R, V-I, J-K, and V-K colors from \cite{Bes88} and AQ00. Specifically, we estimate A$_V$ from four different ratios, each derived from the Savage \& Mathis (1977) reddening law: A$_V$/E(V-R)  =  4.83; A$_V$/E(V-K)  =  1.05; A$_V$/E(J-K)  =  5.80; and A$_V$/E(V-I)  =  2.26. These A$_V$ are collected in Table \ref{tbl-AV}. Colors and spectral types are inconsistent with a field-wide average $<$A$_V>$ for the $\delta$ Cep field. The spatial distribution of the average reddening star to star is shown in Figure \ref{fig-1}. A simple uniform extinction would predict a correlation between A$_V$ and distance, with more distant objects having higher A$_V$. This correlation is absent in Table~\ref{tbl-SPP}, suggesting that either the extinction or the distances are in error. Alternatively, a patchy distribution of the ISM would destroy any correlation, a distinct possibility for this field at Galactic latitude, l  =  0\fdg5. As we shall see, the reddening for reference star DC-2 (discussed in Section~\ref{AbsPi}),  is of critical importance to an estimate of the reddening at the location of $\delta$ Cep.

\subsection{Reference Frame Absolute Parallaxes} \label{Prior}

We have prior knowledge that reference star DC-2 is thought to be physically associated with $\delta$ Cep. \cite{Hof82} ( =  BSC82) note common proper motion with $\delta$ Cep and that DC-2  =  ADS15987C ($\delta$ Cep  =  ADS15987A), while \nocite{deZ99} de Zeeuw et al. (1999) include both $\delta$ Cep and DC-2 in the newly discovered Cep OB6 association. Consequently, we first explored the astrometric properties of the four remaining reference stars, solving for parallax and proper motion of each in turn relative to a reference frame defined by the other three reference stars. We obtained no significant improvement in $\chi^2$ by allowing any reference star (other than DC-2) to have a proper motion relative to the other three. In each case the spectrophotometric parallaxes discussed below entered the solution as observations. 

We derive absolute parallaxes with M$_V$ values from AQ00 and the $<$A$_V> $ obtained from the photometry. These are listed in Table \ref{tbl-SPP}, with three possible values for reference star DC-2, two depending on past spectral classifications. The last value for DC-2 is derived by constraining it to belong to the Cep OB6 association (see section~\ref{AbsPi} below).
The weighted average absolute parallax for the reference frame is $<\pi_{abs}> =  0.77$ mas, including the highest weight parallax determination for DC-2, and 0.63 mas without DC-2. Statistically, DC-2 has very little weight in our reference frame. Nonetheless, it is astrometrically critical, as discussed in (Section~\ref{AbsPi}), below.

\section{Absolute Parallax of $\delta$ Cep}
\subsection{The Astrometric Model}

With the positions measured by FGS 3 we determine the scale, rotation, and offset ``plate
constants" relative to an arbitrarily adopted constraint epoch (the so-called ``master plate") for
each observation set (the data acquired at each epoch). The mJD of each observation set is listed in Table~\ref{tbl-LOO}, along with the magnitude measured by the FGS (zero-point provided by Barnes et al. 1997\nocite{Bar97}), a phase (based on P  =  5.366316 days), and a B-V estimated by comparison with the UBV photometry of \cite{Bar97}. The $\delta$ Cep reference frame contains 5 stars. We employ the six parameter model discussed in McArthur et al. (2001) for those observations. For the $\delta$ Cep field four of the reference stars are significantly redder than the science target and one is bluer. Hence, we apply the corrections for lateral color discussed in Benedict et al. (1999). 

As in all our previous astrometric analyses, we employ GaussFit (\cite{Jef87}) to minimize $\chi^2$. The solved equations
of condition for $\delta$ Cep are:
\beq
        x'  =  x + lcx(\it B-V) - \Delta XFx
\eeq
\beq
        y'  =  y + lcy(\it B-V) - \Delta XFy
\eeq
\beq
\xi  =  Ax' + By' + C + R_x (x'^2 + y'^2) - \mu_x \Delta t  - P_\alpha\pi_x
\eeq
\beq
\eta  =  -Bx' + Ay' + F + R_y(x'^2 + y'^2) - \mu_y \Delta t  - P_\delta\pi_y
\eeq
where $\it x$ and $\it y$ are the measured coordinates from {\it HST};
$\it lcx$ and $\it lcy$ are the
lateral color corrections from Benedict et al. 1999\nocite{Ben99}; and $\it B-V $ are
the  B-V  colors of each star, including the variable B-V of $\delta$ Cep (Table~\ref{tbl-LOO}). Here $\Delta$XFx and $\Delta$XFy are the cross filter corrections in $\it x$ and $\it y$, applied to the observations of $\delta$ Cep and reference star DC-2. $\delta$ Cep has a full range of  0.2 $<$ B-V $<$ 0.6. For this analysis we linearly interpolate between the 1995 and 1998 cross filter calibrations (see Table 1, Benedict et al. 2002) as a function of $\delta$ Cep color.  A  and  B   
are scale and rotation plate constants, C and F are
offsets; $R_x$ and $R_y$ are radial terms;
$\mu_x$ and $\mu_y$ are proper motions; $\Delta$t is the epoch difference from the mean epoch;
$P_\alpha$ and $P_\delta$ are parallax factors;  and $\it \pi_x$ and $\it \pi_y$
 are  the parallaxes in RA and Dec. We obtain the parallax factors from a JPL Earth orbit predictor (\cite{Sta90}), upgraded to version DE405. Orientation to the sky is obtained from ground-based astrometry 
(USNO-A2.0 catalog, Monet 1998\nocite{Mon98}) with uncertainties in the field orientation $ \pm  0\fdg05$.
 
\subsection{Assessing Reference Frame Residuals}
\begin{figure}
\epsscale{0.6}
\plotone{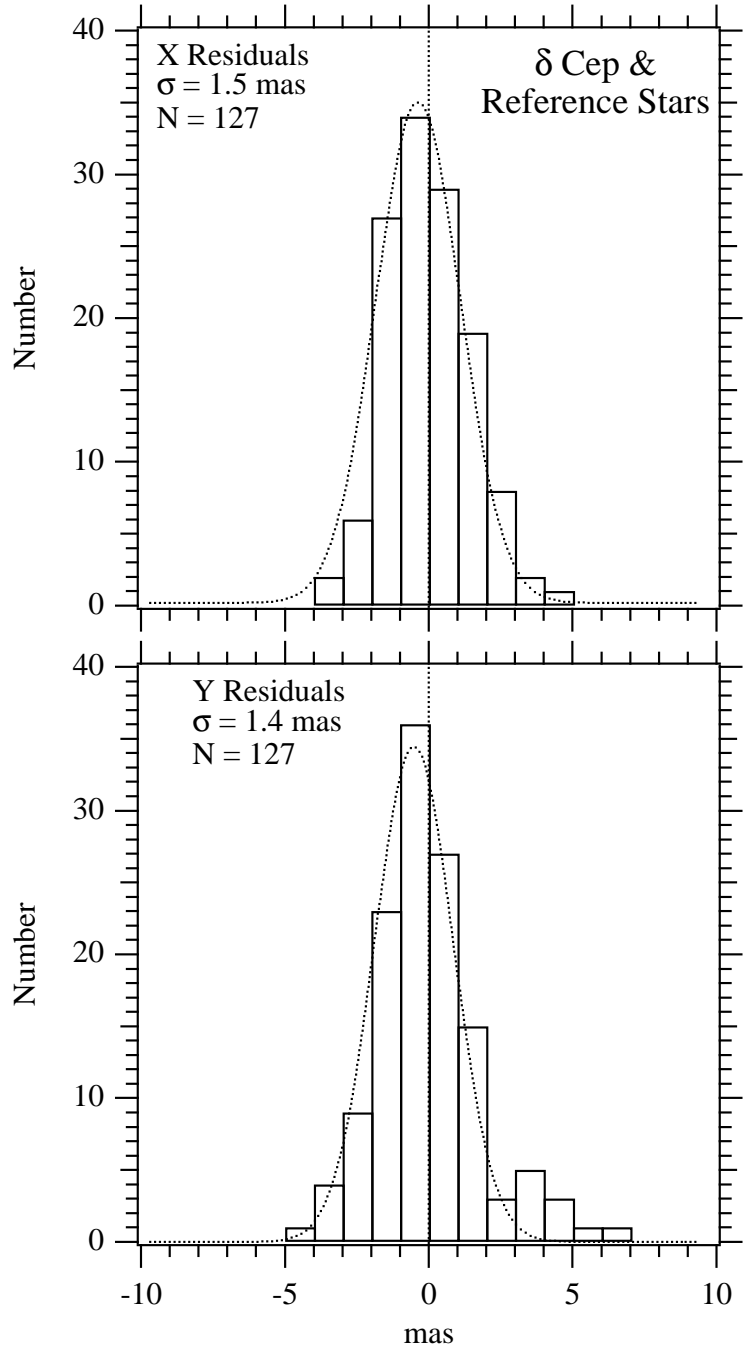}
\caption{Histograms of x and y residuals obtained from modeling $\delta$ Cep and the astrometric reference stars with equations 4 and 5. Distributions are fit
with Gaussians whose $\sigma$'s are noted in the plots.} \label{fig-4}
\end{figure}

The Optical Field Angle Distortion calibration (\cite{McA97}) reduces as-built {\it HST} telescope and FGS 3 distortions with magnitude $\sim1\arcsec$ to below 2 mas  over much of the FGS 3 field of regard. From histograms of the astrometric residuals (Figure~\ref{fig-4}) we conclude that we have obtained correction at the $\sim 1.5$ mas level. The resulting reference frame `catalog' in $\xi$ and $\eta$ standard coordinates (Table \ref{tbl-POS}) was determined
with	$<\sigma_\xi> =  0.3$	 and	$<\sigma_\eta>  =  0.3$ mas.

Noting that the residual histograms have larger dispersions than we typically achieve, we plotted the $\delta$ Cep reference frame residuals against a number of spacecraft, instrumental, and astronomical parameters to determine if there might be unmodeled - but possibly correctable -  systematic effects at the 1 mas level. The plots against residual included $\it x$ and $\it y$ position within the pickle; radial distance from the pickle center; reference star V magnitude and B-V color; and epoch of observation.  Except for reference star DC-2 ( = HD 213307) discussed below, we saw no obvious trends, other than an expected increase in positional uncertainty with reference star magnitude. The largest residuals are associated with observations of reference stars DC-5 and DC-7 made during the two orbits with anomalous drift, discussed in Section~\ref{AstRefs}.

\subsection{A New Companion for HD 213307?}
BSC82 notes a possible very short period companion (P$<$1$^d$) to reference star DC-2  =  HD 213307. Such a companion would be undetectable by the FGS, either directly (changes in fringe structure) or indirectly (astrometric perturbation of the primary). Nonetheless, for DC-2 we found clear long-term and non-linear trends in the residuals with time. Because {\it HST} provides only relative proper motions, we do not expect full agreement with the {\it HIPPARCOS} absolute proper motions. However, the agreement between {\it HIPPARCOS} and {\it HST} for $\delta$ Cep was within the errors, while that for DC-2 was not (Table~\ref{tbl-SUM}). To assess the possibility that the residuals are caused by a perturbation due to a longer-period unseen companion we employed the model from our past binary star work (Benedict et al. 2001),

\beq
\xi  =  Ax' + By' + C + R_x (x'^2 + y'^2) - \mu_x \Delta t  - P_\alpha\pi_x - ORBIT_x
\eeq
\beq
\eta  =  -Bx' + Ay' + F + R_y(x'^2 + y'^2) - \mu_y \Delta t  - P_\delta\pi_y - ORBIT_y
\eeq

\noindent where ORBIT is a function of the traditional astrometric orbital elements. Due to the small number of epochs and observations we were unable to treat HD 213307 as we did Wolf 1062A (Benedict et al. 2001) and simultaneously solve for all the terms in equations 6 and 7. To obtain a solution we constrained all but the parallax, proper motion, and ORBIT parameters to values previously determined using equations 4 and 5. The resulting orbital elements in Table~\ref{tbl-DC2} should be taken as highly preliminary. Figure~\ref{fig-6} shows the residuals from the solution provided by equations 6 and 7 and the fitted orbit. Introducing these parameters to the modeling process reduced the residual histogram dispersions seen in Figure~\ref{fig-4} from $\sigma_x  =  1.5$ to $\sigma_x  =  1.4$ mas and from $\sigma_y  =  1.4$ mas to
$\sigma_y  =  1.0$ mas.

With $M =  4 M_{\sun}$ adopted as the mass of the B7-8 primary (AQ00), the perturbation orbit size, $\alpha  =  2.0$ mas, and the period, P  =  1.06 years yield a component B mass, $M_B\sim1.6M_{\sun}$. If the companion is on the main sequence, this F0 V star would be approximately 7 mas (1.9 AU) distant from and $\sim$3.8 mag fainter than the primary, consistent with the absence of a previous detection. Radial velocity variations of the primary would have an amplitude, K1 $\sim$ 15 \kms, difficult to detect in a B7-8 III/IV star with $vsini$ = 135 \kms (BSC82).

\subsection{The Absolute Parallax of $\delta$ Cep and HD 213307} \label{AbsPi}
In a quasi-Bayesian approach the lateral color and cross-filter calibration values were entered into the model as observations with associated errors.
The reference star spectrophotometric absolute parallaxes also were input as observations with associated errors, not as hardwired quantities known to infinite precision. This approach allows us to incorporate any measurements relevant to our investigation. These include the {\it HIPPARCOS} parallaxes and proper motions of $\delta$ Cep and HD 213307 (HIP 110988) with errors. 

Even though DC-2 is a binary with a poorly determined orbit, we find that we must include this reference star in the solution. An ideal astrometric reference frame would surround $\delta$ Cep. From Figure~\ref{fig-1} it is clear that DC-2 is necessary to minimize extrapolation of the scale determined by the bulk of the offset reference frame. Without DC-2 the reference frame geometry is even less ideal. DC-2 also provides a constraint on the cross filter calibration, because both DC-2 and $\delta$ Cep were observed with the neutral density filter. 
Following a suggestion from the referee, we constrained the difference in parallax between $\delta$ Cep and DC-2, using our prior knowledge of their proximity (see Section~\ref{Prior}). From \cite{deZ99} figure 25 we can estimate that
the 1$\sigma$ dispersion in Galactic longitude for the OB association thought to contain both $\delta$ Cep and DC-2 is
3\arcdeg.  One can therefore infer that the 1$\sigma$ dispersion in
distance in this group is 3\arcdeg/radian $\sim $5\%.  Hence, the 1$\sigma$
dispersion in the parallax difference between two group
members (e.g. DC-2 and $\delta$ Cep) is
\beq
\Delta \pi  =  5\% \times \sqrt{2} \times 3.7 ~mas  =  0.26 ~mas
\eeq
where we have here adopted the mean parallax of Cep OB6, $<\pi> =  3.7$ mas, from \cite{deZ99}.
This estimated difference allows us to assign a higher statistical weight to reference star DC-2 than if we had adopted either past spectral typing (lines 1 and 2 in Table~\ref{tbl-SPP}) or the {\it HIPPARCOS} value. The parallax difference between $\delta$ Cep and DC-2 becomes an observation with associated error fed to our model, an observation used to estimate the parallax difference between the two stars, while solving for the parallax of $\delta$ Cep. 

We obtain for $\delta$ Cep an absolute parallax, $\pi_{abs}  =  3.66  \pm 0.15$ mas, and for DC-2, $\pi_{abs}  =  3.65  \pm 0.15$. Introducing the Cep OB6 parallax dispersion constraint and the {\it HIPPARCOS} parallax and proper motion measurements with their associated errors allows us to obtain a statistically significant result from a reference frame with very poor geometry.
Our final $\delta$ Cep parallax differs by $\sim1\sigma_{\it HIP}$ and by $\sim3\sigma_{\it HST}$ from that measured by {\it HIPPARCOS}, $\pi_{abs}  =  3.32  \pm 0.58$ mas. 
Nordgren et al. (2002) have used long-baseline interferometry to measure the angular diameter of $\delta$ Cep and other Cepheids. Their calibration of the Barnes-Evans (1976) relationship (surface brightness vs color index) yields a distance to $\delta$ Cep of 272 $\pm$  6 pc and a parallax $\pi_{abs}  =  3.68 \pm 0.08$ mas. We note that our $\delta$ Cep absolute parallax and that of Nordgren et al. agree within each other's errors.

\begin{figure}
\epsscale{0.8}
\plotone{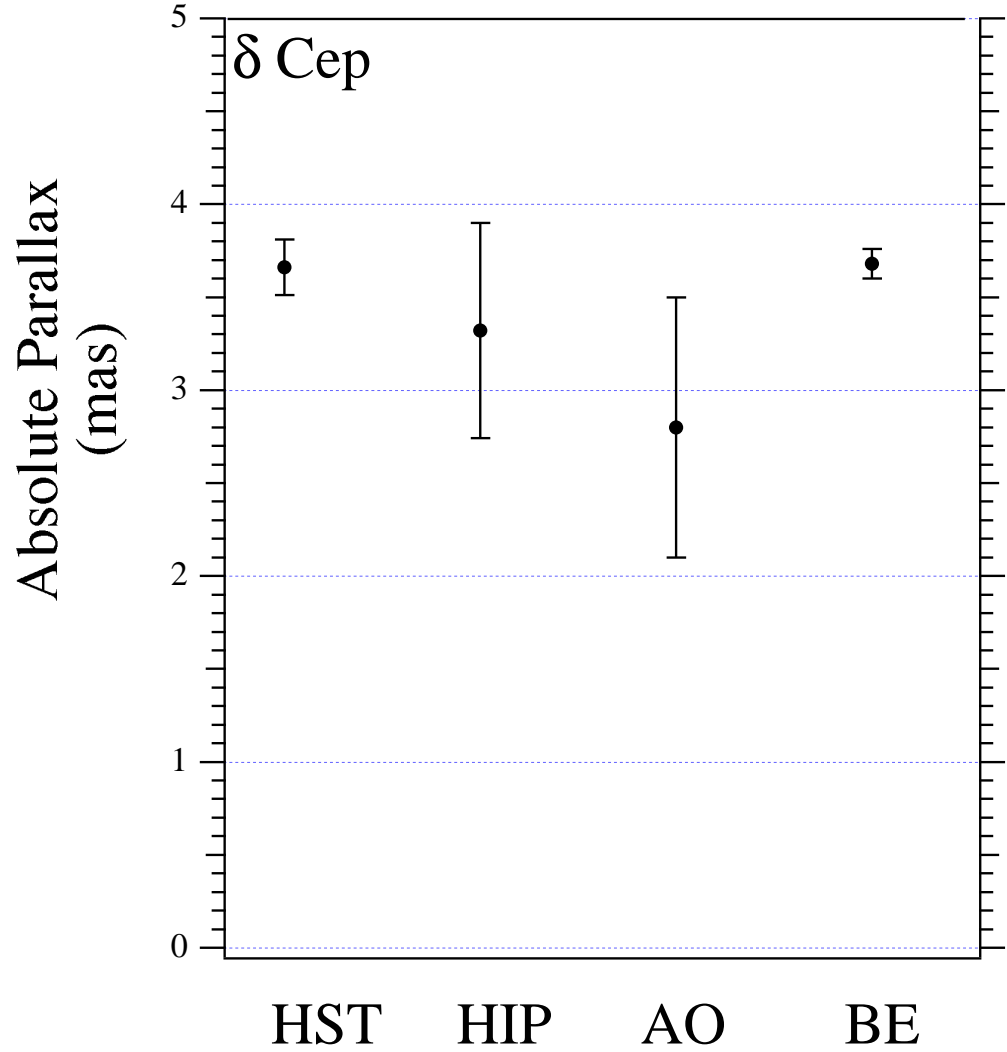}
\caption{ Absolute parallax determinations for $\delta$ Cep. We compare astrometric results from
{\it HST}, {\it HIPPARCOS}, and a recent determination from Allegheny Observatory (AO, Gatewood et al. 1998). Nordgren et al. (2002) have derived a parallax from a new calibration of the Barnes-Evans relation, denoted BE. } 
\label{fig-5}
\end{figure}

\begin{figure}
\epsscale{0.85}
\plotone{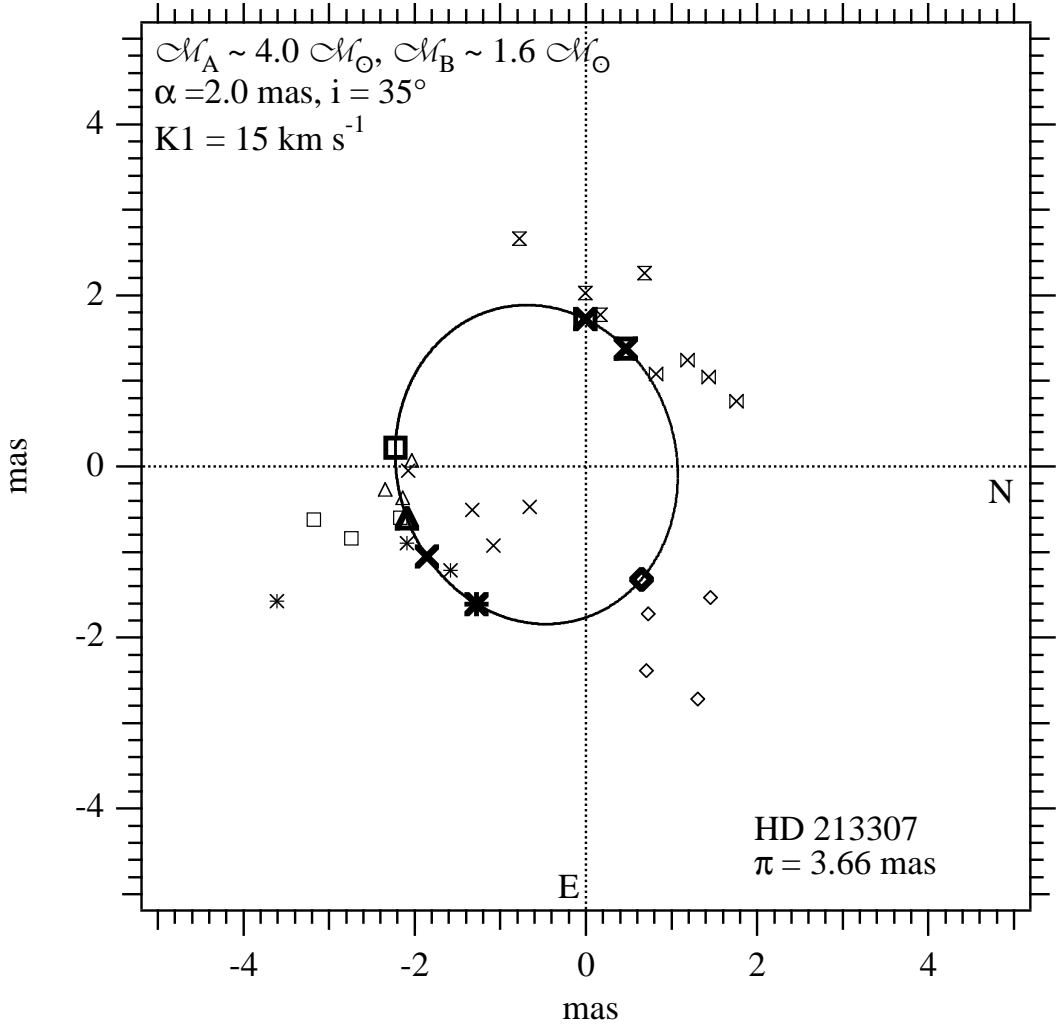}
\caption{Reference star DC-2 (HD 213307) astrometric residuals from modeling with equations 6 and 7. Each observational epoch (denoted Data Set in Table~\ref{tbl-LOO}) is coded with a unique symbol. The emphasized symbols are plotted on the derived orbit at the epoch of observation. The orbit is preliminary, and suggests a $M  =  1.6 M_{\sun}$ companion.} \label{fig-6}
\end{figure}

In Figure \ref{fig-5} we compare $\delta$ Cep astrometric parallax results from {\it HST}, {\it HIPPARCOS}, and Allegheny Observatory (AO). We plot the AO re-determination reported in Gatewood et al. (1998), not the original result (Gatewood et al. 1993). Also plotted is the Nordgren et al. (2002) result. Parallax and proper motion results from {\it HST}, {\it HIPPARCOS}, and AO are collected in Table~\ref{tbl-SUM}. Parallax and proper motion results for reference star DC-2 are also presented in Table~\ref{tbl-SUM}, because it was measured by {\it HIPPARCOS}.

\section{Discussion and Summary}
\subsection{{\it HST} Parallax Accuracy}
\begin{figure}
\epsscale{0.85}
\plotone{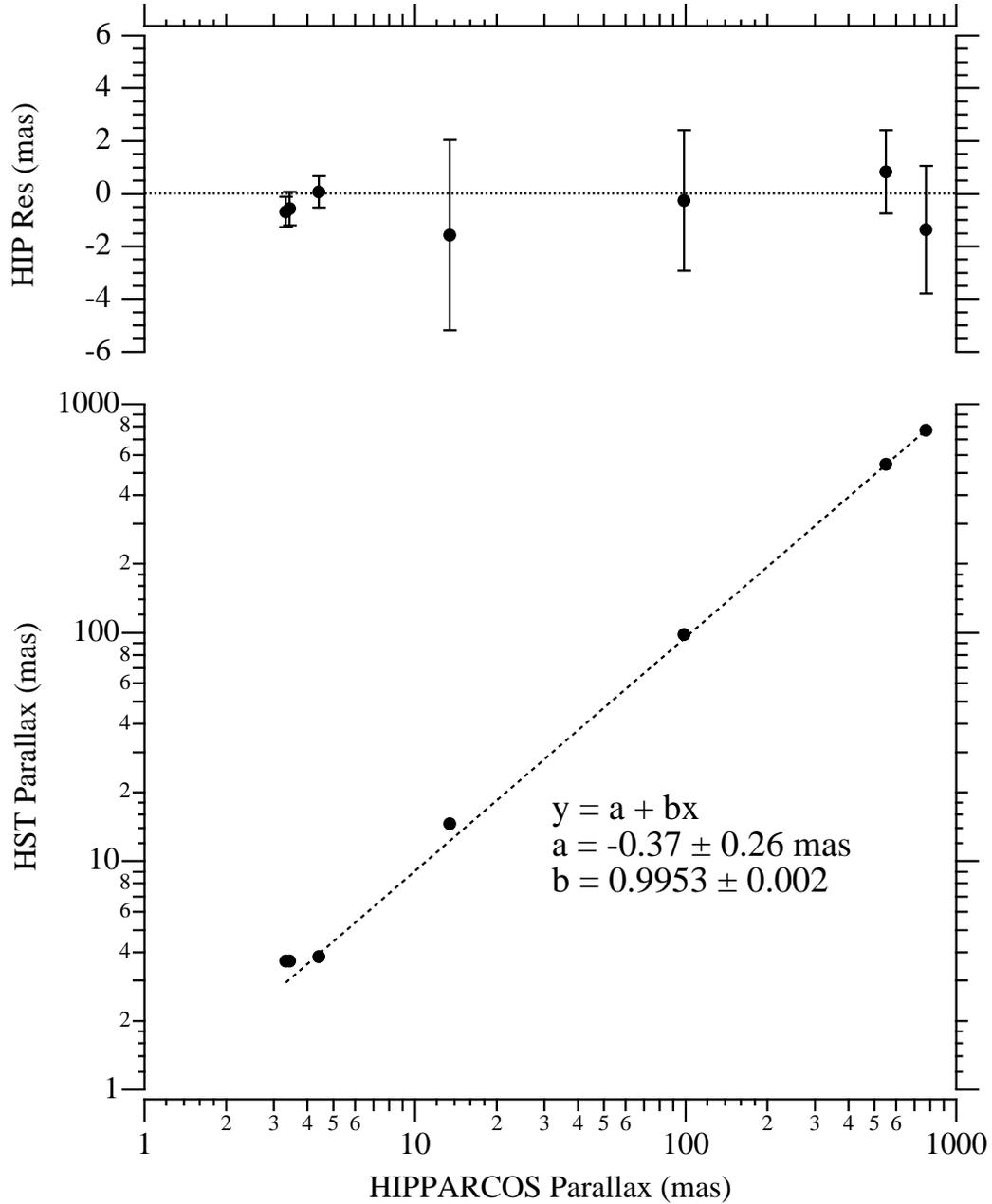}
\caption{Bottom: {\it HST} absolute parallax determinations compared
with {\it HIPPARCOS} for the seven targets listed in Table~\ref{tbl-HH}. Top: The {\it HIPPARCOS} residuals to the error-weighted impartial regression line. The error bars on the residuals are {\it HIPPARCOS} 1$\sigma$ errors. 
} \label{fig-7}
\end{figure}

Our parallax precision, an indication of our internal, random error, is often less than 0.4 mas. To assess our accuracy, or external error, we must compare our parallaxes with results from independent measurements. Following \cite{Gat98}, we plot all parallaxes obtained by the {\it HST} Astrometry Science Team with FGS 3 against those obtained by {\it HIPPARCOS}. Data for these seven objects are collected in Table \ref{tbl-HH} and shown in Figure \ref{fig-7}. We have not included four Hyades stars whose parallaxes are considered preliminary (van Altena et al. 1997). The dashed line is a weighted regression that takes into account errors in both input data sets. 

The regression indicates a 2.5$\sigma$ scale difference between the {\it HIPPARCOS} and {\it HST} results. In addition, the fit $\chi^2$ suggests that either the {\it HIPPARCOS} or {\it HST} errors are overstated. Some, but not all of this discrepancy can be explained by pointing out that our log-log plot artificially spreads out these data which are effectively in two clumps, large and small parallaxes.
Further exploring this issue, we have conducted a fully Bayesian analysis of the data in question, with an additional factor that represents the degree to which the presumed errors agree with the fit. We find that the input standard deviations of the {\it HST} and {\it HIPPARCOS} data points may have been overstated by a factor of $\sim1.5$. It is not possible to decide which errors, {\it HST} or the {\it HIPPARCOS}, or some combination of the two, are overstated. However, {\it HIPPARCOS} errors have been subjected to many tests confirming their validity. Hence, it is likely that {\it HST} errors are overstated. 

Regarding the scale difference, we compare two hypotheses; the null hypothesis (a = 0, b = 1) and the complex hypothesis (a $\neq$ 0, b $\neq$ 1). We find that the Bayes factor against the null hypothesis (that the straight line in Figure~\ref{fig-7} is correctly described by a = 0, b = 1) and in favor of a more complex hypothesis is about 30, depending on the priors we put on a and b. These 30:1 odds provide only modest support for the complex hypothesis, that there is a systematic scale deviation between the {\it HST} and {\it HIPPARCOS} results. We point out that the amount of data is very small, and that much of the scale difference (if real) depends upon the two largest parallaxes, which have a great deal of leverage. 

Measured proper motions provide another argument against the reality of this scale difference. Because it is desirable to reduce the impact of proper motion errors on an {\it HST} -- {\it HIPPARCOS} comparison, we consider only two of the objects in Table~\ref{tbl-HH} (Proxima Cen and Barnard's Star). They have proper motion vector lengths exceeding 3800 mas yr$^{-1}$. {\it HST} yields proper motions relative to a local reference frame and {\it HIPPARCOS} produced absolute proper motions. The dissimilar approaches can result in unequal proper motions for the same object. However, comparing {\it HST} with {\it HIPPARCOS}, the average difference between these proper motion vectors is -0.01\%, indicating a negligible scale difference. 

\subsection{The Lutz-Kelker Bias}

When using a trigonometric parallax to estimate the absolute
magnitude of a star, a correction should be made for the
Lutz-Kelker (LK) bias (\cite{Lut73}).
Because of the galactic latitude and distance of $\delta$ Cep, 
and the scale height of the
stellar population of which it is a member,
we  use a uniform space density  for calculating
the LK bias.
An LK
algorithm modified by Hanson (1979) that includes the power law
of the
parent population is used. A correction of -0.015  $\pm$   0.01 mag is derived for
the
LKH bias correction to the {\it HST} parallax of $\delta$ Cep. The LKH bias is small because $\sigma_{\pi} \over \pi$$\sim$ 4\% is small.

\subsection{The Absolute Magnitude of $\delta$ Cep} \label{AV1}
To obtain the intrinsic absolute magnitude of $\delta$ Cep we require the intensity-averaged apparent magnitude, the absolute parallax, and an estimate of interstellar extinction. To estimate the extinction we turn to reference star DC-2 (HD 213307). For this star we obtain an absolute parallax, $\pi_{abs}  =  3.65$ mas, between the two estimates depending on past spectral type and luminosity class determinations (Table~\ref{tbl-SPP}). From SIMBAD we collected B-V for all B8 III and B7 IV stars with V $\leq$ 4 (presumably the nearest and least reddened members of their respective classes). The average for B8 III was $<$B-V$>$  =  -0.09 and for B7 IV $<$B-V$>$   =  -0.13. We adopt an intermediate (B-V)$_0$  =  -0.11. From the measured color, B-V  =  -0.04, we obtain a color excess, E(B-V) =  0.07. Because a blue star can yield a different ratio of total to selective absorbtion, R, than a redder star, depending on the amount of color excess, we employ the formulation (Laney \& Stobie 1993), 
\beq
R  =  3.07 + 0.28\times (B-V)_0 + 0.04\times E(B-V)
\eeq
and obtain for DC-2 $<$ A$_V>$  =  0.21 $\pm$ 0.03. The absolute magnitude of HD 213307 is then M$_V$  =  -1.10. Given that the distance to $\delta$ Cep and DC-2 are the same within the errors, we adopt the same color excess, E(B-V) =  0.07 for $\delta$ Cep, yielding (B-V)$_0$  =  0.59 and R  =  3.24. Therefore, $<$ A$_V>$  =  0.23 $\pm$  0.03 for $\delta$ Cep. Our determination agrees with $<$ A$_V>$  =  0.25 $\pm$  0.06 from Fernie et al. (1995)\footnote{http://ddo.astro.utoronto.ca/cepheids.html}.  Because differential reddening effects are smaller at the I-band, we use a Savage \& Mathis R  =  3.1 reddening law to provide A$_I$ =  0.14.

We next require the intensity-averaged magnitude of $\delta$ Cep. From \cite{Fea97} we obtain an intensity weighted average $<$V$>$ =  3.954. From the photometry of Moffett \& Barnes (1985), we adopt a Johnson I-band intensity averaged magnitude, $<$I$>$ =  2.993. With the {\it HST} absolute parallax from Section \ref{AbsPi} and  $<$ A$_V>$  =  0.23 $\pm$  0.03, we derive M$_V  =  -3.47 \pm 0.10$, including the LKH bias and associated uncertainty. Similarly, M$_{I(CK)}  =  -4.14 \pm 0.10$, where we have transformed the Johnson to the Cousins-Kron photometric system as per Bessell (1979). The intensity-averaged photometry and absolute magnitudes are collected into Table~\ref{tbl-CepAbsMag}.
Using surface-brightness techniques, Gieren et al. (1993) obtain for $\delta$ Cep an absolute visual magnitude M$_V  =  -3.63 \pm 0.13$ and, for log P  =  0.73, M$_V  =  -3.55 \pm 0.11$ from their Period-Luminosity (P-L) calibration for 100 Cepheids. Using a large number of relatively imprecise HIPPARCOS parallaxes, \cite{Fea97} and \cite{Lanoix99} derive slightly different P-L calibrations for Cepheid variable stars. For log P  =  0.73, Feast obtains M$_V$ =  -3.48 and Lanoix et al., M$_V$ =  -3.46. All determinations agree with our value within the errors. 

\subsection{The Distance Modulus of the LMC} \label{LMCDM}
Uncertainty in the distance to the Large Magellanic Cloud (LMC) contributes  a substantial fraction of the uncertainty in
the Hubble Constant (\cite{Mould00}).
The HST Key Project on the Extragalactic Distance Scale (\cite{Freed01},
\cite{Mould00}) 
and the Type Ia Supernovae Calibration Team (\cite{Saha99}) have adopted
the distance modulus value m-M  =  18.5.  Values from 18.1 to 18.8 are reported in the current
literature, with those less than 18.5 supporting the short distance scale and 
those greater than 18.5, the long distance scale.
Comprehensive reviews of the methods can be found in
\nocite{Car00}Carretta \etal (2000), \cite{Gib99}, and \cite{Cole98}. 

Ideally, our absolute magnitude values for $\delta$ Cep, M$_V  =  
-3.47 \pm 0.10$ and M$_{I(CK)}  =  -4.14 \pm 0.10$, combined with apparent magnitudes for LMC Cepheids (corrected for reddening internal to the LMC) at log P  = 0.73 would yield a unique LMC distance modulus. However, there are several complications that render such  a distance modulus suspect. The first of these is cosmic dispersion in Cepheid properties due to the finite width of the instability strip. As shown in figure 5.13 of Binney \& Merrifield (1998), lines of constant period are not horizontal in the HR diagram. This complication motivates the determination of Period-Luminosity-Color (P-L-C) relationships. Udalski et al. (1999) \nocite{Uda99} find a dispersion $\sigma$ = 0.07 magnitude about the P-L-C relation for LMC Cepheids. The scatter about any P-L relation decreases markedly as one utilizes progressively redder band passes that are less affected by internal reddening (\cite{Uda99} and \cite{Tan99}). A reddening-free magnitude, W, yields the P-L relationship with the least dispersion, \nocite{Uda99} Udalski et al. (1999) finding $\sigma$ = 0.08 mag. Assuming that the remaining scatter in a P-L-C or reddening-free P-L relationship is cosmic scatter, we adopt 0.075 mag as the cosmic dispersion in the intrinsic properties of Cepheid variables, including the defining member of the class, $\delta$ Cep. 

Varying amounts of line blanketing due to intrinsic metallicity differences can also affect the apparent magnitude of a star. A comprehensive discussion of its effect on the Cepheid P-L relation is presented by \cite{Freed01}. We adopt their correction, used for both V and I band and differential with respect to the LMC, of -0.2 $\pm$  0.2 mag dex$^{-1}$.  Specific to our calibration of the LMC P-L relation, for $\delta$ Cep \cite{And02} find [O/H]  =  +0.01. The LMC has [O/H] $\sim$ -0.4 (\cite{Ken98}). Hence, if $\delta$ Cep had LMC metallicity, it would be 0.08 $\pm$  0.08 mag brighter in V and I. 

We adopt the OGLE LMC V and I P-L relations (\cite{Uda99}) because they are based on a very large number of Cepheids. Utilizing these to generate apparent, absorption-corrected Cepheid magnitudes at log P  =  0.73, we obtain LMC V-band distance moduli m-M  =  18.50  $\pm$  0.13 or, corrected for metallicity, m-M  =  18.58 $\pm$  0.15. The corresponding I-band moduli are m-M  =  18.53  $\pm$  0.13 and m-M  =  18.61 $\pm$  0.15, respectively. The errors include an RSS-ed 0.075 magnitude cosmic dispersion. We list these distance moduli in Table~\ref{tbl-CepLMC}.
Distance moduli corrected for metallicity are listed in the rightmost column as m-M = f(Z). At this stage in its maturity, this term introduces as much uncertainty as correction.

\begin{figure}
\epsscale{0.7}
\plotone{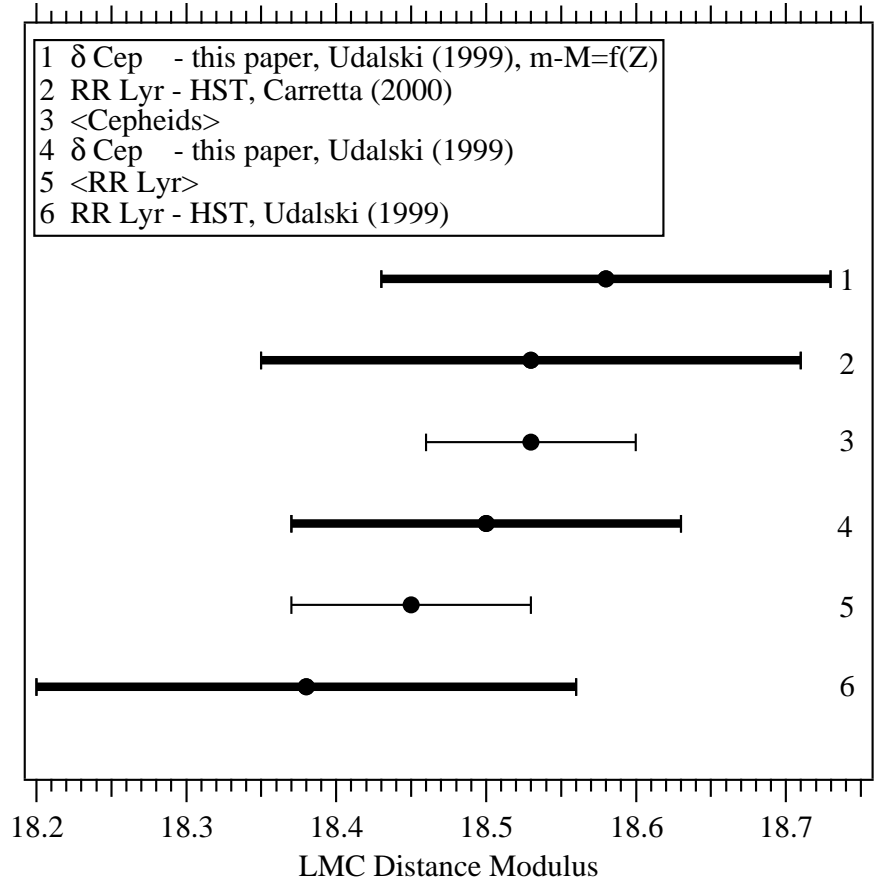}
\caption{Recent determinations of the V-band distance modulus of the Large Magellanic 
Cloud (see Table~\ref{tbl-LMC}). Values labeled 3 and 5 are weighted averages of a number of independent determinations using Cepheids (3) and RR Lyr variables (5).
} \label{fig-8}

\end{figure}

All estimates agree within their respective errors with our recent determinations from {\it HST} astrometry of RR Lyr (\cite{Ben02}). There we obtained values of 18.53  $\pm$   0.10 and 18.38  $\pm$   0.10, where the range comes from differing $<$V$_0$(RR)$>$ values for LMC RR Lyr variables found in the literature. We note that the errors associated with the Benedict et al. (2001) LMC distance moduli derived from RR Lyr neglected cosmic dispersion. According to \cite{Pop98} this amounts to $\sim0.14$ magnitudes, increasing our distance moduli errors to 0.18 magnitude. Table~\ref{tbl-LMC} and Figure~\ref{fig-8} summarize LMC distance modulus determinations based on {\it HST} astrometry and compare them with weighted averages of a number of results based on RR Lyr and Cepheid variables. A more comprehensive version of this figure, along with the corresponding complete table and references, summarizing over 80 determinations
based on 21 independent methods, can be found on the web\footnote{http://clyde.as.utexas.edu/SpAstNEW/head.ps}.  All LMC distance moduli based on {\it HST} astrometry are consistent with the LMC m-M = 18.50 $\pm$  0.10 value adopted by the {\it HST} Distance Scale Key Project (\cite{Freed01}).

\subsection{Summary}
{\it HST} astrometry yields an absolute trigonometric parallax for $\delta$ Cep, $\pi_{abs}  =  3.66  \pm 0.15$ mas. This high-precision result requires an extremely small Lutz-Kelker bias correction, -0.015 $\pm$  0.01 magnitude. To reduce our astrometric residuals to near-typical levels requires that we model reference star DC-2 as a binary and constrain it and $\delta$ Cep to belong to the same stellar group, Cep OB6. Our astrometric results for DC-2 yield an extinction for that star of $<$ A$_V>$  =  0.21 $\pm$  0.03. Correcting for color-dependent R differences, we find $<$ A$_V>$  =  0.23 $\pm$  0.03 for $\delta$ Cep. The dominant contributor to the error in the resulting absolute magnitude for $\delta$ Cep, M$_V  =  -3.47  \pm  0.10$,  remains the parallax. We find an LMC V-band distance modulus m-M  =  18.50  $\pm$  0.13, uncorrected for metallicity. This value is in agreement with our previous determinations with {\it HST} astrometry of RR Lyr and the value adopted by the {\it HST} Distance Scale Key Project (\cite{Freed01}).

\acknowledgments

Support for this work was provided by NASA through grants GTO NAG5-1603 from the Space Telescope 
Science Institute, which is operated
by the Association of Universities for Research in Astronomy, Inc., under
NASA contract NAS5-26555. These results are based partially on observations obtained with the
Apache Point Observatory 3.5m telescope, which is owned and operated by
the Astrophysical Research Consortium. This publication makes use of data products from the Two Micron All Sky Survey,
which is a joint project of the University of Massachusetts and the Infrared Processing
and Analysis Center/California Institute of Technology, funded by the National
Aeronautics and Space Administration and the National Science Foundation. 
This research has made use of the SIMBAD database, operated at CDS, Strasbourg, France; the NASA/IPAC Extragalactic Database (NED) which is operated by the
Jet Propulsion Laboratory, California Institute of Technology, under contract with the National Aeronautics and Space Administration;  and NASA's Astrophysics
Data System Abstract Service. 
Thanks to Tom Barnes for helpful discussions and an early review of the text. Thanks to the many people in Danbury CT who have surfed through many waves of change (once Perkin-Elmer, then Hughes Aerospace, then Raytheon, now Goodrich) and continue to support the FGS, especially Linda Abramowicz-Reed. Finally, we thank Andy Gould for his careful and critical refereeing, and for his suggestion to relate the reference star DC-2 and $\delta$ Cep parallaxes to each other through their membership in Cep OB6.
\clearpage



\begin{center}
\begin{deluxetable}{llllll}
\tablewidth{0in}
\tablecaption{Astrometric Reference Stars: Photometry \label{tbl-VIS}}
\tablehead{\colhead{ID}&
\colhead{V} &\colhead{B-V} &
\colhead{V-R} &
\colhead{V-I} &
\colhead{V-K} }
\startdata
DC-2\tablenotemark{a}&6.30$\pm0.02$&-0.04$\pm0.02$&&&\nl
DC-3&13.47$\pm0.02$&1.60$\pm0.10$&1.10$\pm0.04$&2.17$\pm0.04$&4.22$\pm0.06$\\
DC-4&12.68$\pm0.02$&1.83$\pm0.10$&1.34$\pm0.03$&2.56$\pm0.03$&4.93$\pm0.06$\\
DC-5&13.68$\pm0.02$&1.34$\pm0.10$&0.96$\pm0.04$&1.94$\pm0.04$&3.69$\pm0.06$\\
DC-7&14.18$\pm0.02$&1.60$\pm0.10$&1.11$\pm0.04$&2.22$\pm0.04$&4.19$\pm0.07$\\
\enddata
\tablenotetext{a}{\cite{Lut77}}
\end{deluxetable}
\end{center}

\begin{center}
\begin{deluxetable}{llllll}
\tablewidth{0in}
\tablecaption{Astrometric Reference Stars: Near-IR and Washington-DDO Photometry\label{tbl-IR}} 
\tablehead{\colhead{ID}&
\colhead{K} &
\colhead{J-H} &
\colhead{J-K} &
\colhead{M-T$_2$} &
\colhead{M-51} }
\startdata
DC-2&&&&-0.17$\pm0.02$&0.02$\pm0.02$\\
DC-3&9.25$\pm0.02$&0.76$\pm0.02$&0.93$\pm0.03$&2.22$\pm0.02$&-0.07$\pm0.02$\\
DC-4&7.75$\pm0.02$&0.93$\pm0.02$&1.21$\pm0.03$&2.58$\pm0.01$&-0.1$\pm0.01$\\
DC-5&9.99$\pm0.02$&0.71$\pm0.02$&0.84$\pm0.03$&1.97$\pm0.02$&0.01$\pm0.03$\\
DC-7&9.99$\pm0.02$&0.67$\pm0.02$&0.88$\pm0.03$&2.25$\pm0.01$&-0.05$\pm0.02$\\
\enddata
\end{deluxetable}
\end{center}

\begin{center}
\begin{deluxetable}{llllll}
\tablewidth{6in}
\tablecaption{Astrometric Reference Stars: Spectral Classifications and
Spectrophotometric Parallaxes\label{tbl-SPP}
}
\tablehead{\colhead{ID}& 
\colhead{SpT}&
\colhead{V} & 
\colhead{M$_V$} & 
\colhead{A$_V$}  &
\colhead{$\pi_{abs}$ (mas)}}
\startdata
DC-2&B7 IV\tablenotemark{a}&6.30&-0.85$\pm$0.4\tablenotemark{b}&0.28&4.2$\pm$0.8\\
DC-2&B8 III\tablenotemark{c}&6.30&-1.35$\pm$0.4\tablenotemark{b} &0.09&3.2$\pm$0.6\\
DC-2&B7-8 III-IV\tablenotemark{d}&6.30&-1.10$\pm$0.1 &0.21&3.70$\pm$0.26\\
DC-3&K1 III&13.47&0.6$\pm$0.4&1.63&0.6$\pm$0.1\\
DC-7&G8 III&14.18&0.8$\pm$0.4&2.05&0.5$\pm$0.1\\
DC-4&K3 III&12.68&0.3$\pm$0.4&2.07&0.9$\pm$0.2\\
DC-5&G1 III&13.68&0.9$\pm$0.4&2.13&0.7$\pm$0.1\\
\enddata
\tablenotetext{a}{\cite{Lut77}},\tablenotetext{b}{\cite{Weg00}},
\tablenotetext{c}{\cite{Sav85}}
\tablenotetext{d}{from membership in CepOB6 (\cite{deZ99})}
\end{deluxetable}
\end{center}

\begin{center}
\begin{deluxetable}{llllll}
\tablewidth{0in}
\tablecaption{ Astrometric Reference Stars: A$_V$ from Spectrophotometry  \label{tbl-AV}}
\tablehead{  \colhead{ID}&
\colhead{A$_V$(V-I)}&   \colhead{A$_V$(V-R)}&  \colhead{A$_V$(V-K)} &  \colhead{A$_V$(J-K)} &
\colhead{$<A$$_V$$>$\tablenotemark{a} }}
\startdata
DC-3&1.83&1.43&1.80&1.45&1.63$\pm$0.12\\
DC-7&2.35&1.98&2.13&1.74&2.05$\pm$0.15\\
DC-4&2.06&1.90&2.03&2.32&2.07$\pm$0.10\\
DC-5&2.37&2.02&1.96&2.15&2.13$\pm$0.10\\
$<$A$_V$$>$\tablenotemark{b}&2.15&1.83&1.98&1.91&1.97 $\pm$ 0.13\\
\enddata
\tablenotetext{a}{average by star}
\tablenotetext{b}{average by color index}
\end{deluxetable}
\end{center}

\begin{deluxetable}{lllll}
\tablewidth{4in}
\tablecaption{$\delta$ Cep Log of Observations
\label{tbl-LOO}}
\tablehead{\colhead{Data Set}&
\colhead{mJD} &
\colhead{phase \tablenotemark{a}} &
\colhead{V \tablenotemark{b}} &
\colhead{B-V \tablenotemark{c}} 
}
\startdata
1&49908.60462&0.870&3.967&0.63\\
2&49942.7928&0.241&3.910&0.64\\
3&50087.07986&0.128&3.775&0.55\\
4&50104.13513&0.306&3.989&0.69\\
5&50630.66157&0.423&4.103&0.77\\
6&50668.40227&0.456&4.131&0.79\\
7&50799.52478&0.891&3.862&0.57\\
\enddata
\tablenotetext{a}{Phase based on P  =  5.$^d$366316, T$_0$ =  43673.644 (mJD) (\cite{Bar97})}
\tablenotetext{b}{Differential FGS photometry. Zero point from Barnes et al. (1997)\nocite{Bar97}. }
\tablenotetext{c}{Estimated from phase and Barnes et al. (1997) photometry\nocite{Bar97}.} 
\end{deluxetable}

\begin{center}
\begin{deluxetable}{lllll}
\tablewidth{0in}
\tablecaption{$\delta$ Cep and Reference Stars: Astrometry\label{tbl-POS}}   
\tablehead{\colhead{ID}&
\colhead{$\xi$ \tablenotemark{a}} &
\colhead{$\eta$ \tablenotemark{a}} &
\colhead{$\mu_x$ \tablenotemark{b}} &
\colhead{$\mu_y$ \tablenotemark{b}} }
\startdata
$\delta$ Cep&45.2774$\pm$0.0003&104.9195$\pm$0.0003&0.0174$\pm$0.0002&0.0050$\pm$0.0002\nl
DC-2&37.4174$\pm$0.0003&64.9353$\pm$0.0003&0.0215$\pm$0.0003&0.0037$\pm$0.0003\nl
DC-3\tablenotemark{c}&0.0000$\pm$0.0003&0.0000$\pm$0.0004&&\nl
DC-4&131.7161$\pm$0.0002&116.1802$\pm$0.0003&&\nl
DC-5&88.9765$\pm$0.0003&-27.4270$\pm$0.0003&&\nl
DC-6&167.0683$\pm$0.0003&76.5763$\pm$0.0003&& \nl
\enddata
\tablenotetext{a}{$\xi$ and $\eta$ are positions in arcseconds relative to DC-3
}
\tablenotetext{b}{$\mu_x$ and $\mu_y$ are relative motions in arcsec
yr$^{-1}$ }
\tablenotetext{c}{RA  =  22$^h$ 29$^m$ 04.59$^s$ Dec  =  $+$58\arcdeg 47' 40\farcs7, J2000, epoch  =  mJD
50104.1363}
\end{deluxetable}
\end{center}

\begin{center}
\begin{deluxetable}{ll}
\tablecaption{HD 213307 - Elements of Perturbation Orbit\label{tbl-DC2}
} 
\tablewidth{0in}
\tablehead{\colhead{Parameter} & \colhead{Value} }
\startdata
$\alpha$(mas)& 2.0$\pm$ 0.2\nl
P(days)& 390 $\pm$ 9\nl 
P(years)& 1.07 $\pm$ 0.1\nl
T$_0$ & 2002.3 $\pm$ 0.1 \nl
e& 0.35 $\pm$ 0.09\nl
i(\arcdeg)& 36 $\pm$ 14 \nl
$\Omega$(\arcdeg)&100 $\pm$ 8 \nl
$\omega$(\arcdeg)&259 $\pm$ 7 \nl
\enddata
\end{deluxetable}
\end{center}

\begin{center}
\begin{deluxetable}{lll}
\tablecaption{Final $\delta$ Cep and HD 213307 (DC-2) Absolute Parallax and Relative Proper Motion Compared to Previous Results\label{tbl-SUM}}
\tablewidth{0in}
\tablehead{\colhead{Parameter} &  \colhead{$\delta$ Cep }&  \colhead{HD 213307}}
\startdata
{\it HST} study duration  &2.44 y&\nl
number of observation sets    &   7 &\nl
ref. stars $ <V> $ &  $12.06 $  &\nl
ref. stars $ <B-V> $ &  $1.3 $ &\nl
ref. stars $ <V>$ excl. DC-2  &  $13.50 $ & \nl
ref. stars $ <B-V> $ excl. DC-2&  $1.6 $ &\nl
\nl
{\it HST} Absolute Parallax   & 3.66  $\pm$  0.15 &  3.65  $\pm$  0.15 mas\nl
{\it HIPPARCOS} Absolute Parallax &3.32  $\pm$ 0.58& 3.43  $\pm$ 0.64 mas\nl
AO Absolute Parallax & 2.8  $\pm$  0.7&  mas\nl
\nl
{\it HST} Relative Proper Motion  &17.4  $\pm$  0.7 & 21.8  $\pm$  1.2 mas y$^{-1}$ \nl
 \indent in pos. angle & 73\arcdeg  $\pm$ 3\arcdeg &80\arcdeg  $\pm$ 5\arcdeg\nl
{\it HIPPARCOS} Absolute Proper Motion  &16.9  $\pm$  3.1& 16.1  $\pm$  2.7mas y$^{-1}$ \nl
 \indent in pos. angle & 79\arcdeg  $\pm$ 14\arcdeg & 74\arcdeg  $\pm$ 12\arcdeg\nl
\enddata
\end{deluxetable}
\end{center}

\begin{deluxetable}{llll}
\tablewidth{0in}
\tablecaption{{\it HST} and {\it HIPPARCOS} Absolute Parallaxes    \label{tbl-HH}}
\tablehead{\colhead{Object}& \colhead{{\it HST}}& \colhead{{\it HIP}} &\colhead{{\it HST} Reference}}
\startdata
Prox Cen  &     769.7  $\pm$   0.3 mas  & 772.33 $\pm$   2.42 & Benedict et al. 1999\nl
Barnard's Star& 545.5  $\pm$   0.3   & 549.3  $\pm$   1.58 & Benedict et al. 1999\nl
Feige 24      & 14.6   $\pm$   0.4   & 13.44  $\pm$   3.62 & Benedict et al. 2000\nl
Wolf 1062     & 98.0   $\pm$     0.4   & 98.56  $\pm$   2.66 & Benedict et al. 2001\nl
RR Lyr &3.60   $\pm$   0.20 &   4.38  $\pm$     0.59 & Benedict et al. 2002\nl
$\delta$ Cep &3.66   $\pm$   0.15 &   3.32  $\pm$     0.58 & this paper\nl
HD 213307 &3.65   $\pm$   0.15 &   3.43  $\pm$     0.64 & this paper\nl

\enddata
\end{deluxetable}

\begin{center}        
\begin{deluxetable}{lccc}        
\tablewidth{6in}        
\tablecaption{$\delta$ Cep Intensity-Weighted Magnitudes\label{tbl-CepAbsMag}}  
\tablehead{\colhead{mag } &  \colhead{J\tablenotemark{a}} &  \colhead{CK\tablenotemark{b}}&  \colhead{DDO\tablenotemark{c}}}    
\startdata        
$<V>$&3.95&3.95&3.95\nl
$<I>$&2.99&3.18&\nl
$<V>$-$<I>$&0.96&0.75&\nl
A$_V$&0.23\tablenotemark{d}&&0.25\nl
A$_I$&0.14\tablenotemark{d}&&\nl
$<V_0>$&3.73&3.73&3.70\nl
$<I_0>$&2.84&3.06&\nl
M$_V$&-3.47$\pm$0.10&-3.47$\pm$0.10&\nl
M$_I$&-4.34$\pm$0.10&-4.14$\pm$0.10&\nl
\enddata

\tablenotetext{a} { Johnson VI photometry from Moffett \& Barnes (1985) }
\tablenotetext{b} { Transformation per Bessel (1979). }  
\tablenotetext{c} {from http://ddo.astro.utoronto.ca/cepheids.html}
\tablenotetext{d} {Estimated from Section~\ref{AV1} and Savage \& Mathis (1977) reddening curve.}
\end{deluxetable} 
\end{center}

\begin{center}        
\begin{deluxetable}{llllll}        
\tablewidth{0in}        
\tablecaption{LMC Distance Moduli from the OGLE\tablenotemark{a} Cepheid Period-Luminosity Relation\label{tbl-CepLMC}} 
\tablehead{\colhead{Wavelength} &  \colhead{a}&  \colhead{b}&  \colhead{m} &\colhead{m-M\tablenotemark{b}} & \colhead{m-M = f(Z)\tablenotemark{b,c}}  }  
\startdata
V & -2.760$\pm$0.031 & 17.04$\pm$0.021  & 15.03$\pm$0.03 &18.50$\pm$0.13 &18.58$\pm$0.15\nl 
I & -2.962$\pm$0.021 & 16.56$\pm$0.014  & 14.40$\pm$0.02 &18.53$\pm$0.13 &18.61$\pm$0.15\nl
\enddata
\tablenotetext{a} { Udalski et al. (1999), m  =  a$\cdot$log P + b }, 
\tablenotetext{b} {Errors are RSS of m error from PL, absolute magnitude (M) error, and 0.075 magnitude cosmic dispersion. }, 
\tablenotetext{c} {\cite{Freed01} metallicity correction, $\Delta$(m-M)  =  -0.2$\pm$0.2 mag dex$^{-1}$} 
\end{deluxetable} 
\end{center}   

\begin{center}        
\begin{deluxetable}{llll}        
\tablewidth{6in}        
\tablecaption{ V-band LMC Distance Moduli \label{tbl-LMC}}
\tablehead{\colhead{No.} &  \colhead{Object} &  \colhead{Source}&  \colhead{m-M} }    
\startdata        
1 &      $\delta$ Cep & this paper, m-M = f(Z) & 18.58  $\pm$  0.15\tablenotemark{a} \nl 
2 & RR Lyrae & Benedict et al. 2002 & 18.53  $\pm$  0.18\tablenotemark{b} \nl 
3 &     all Cepheids & Benedict et al. 2002, table 10 & 18.53  $\pm$  0.07\tablenotemark{c} \nl 
4 &      $\delta$ Cep & this paper, m-M$\not = $f(Z) & 18.50  $\pm$  0.13\tablenotemark{d} \nl 
5 &      all RR Lyr & Benedict et al. 2002, table 10 & 18.45  $\pm$  0.08\tablenotemark{e} \nl 
6 &      RR Lyrae & Benedict et al. 2002 & 18.38  $\pm$  0.18\tablenotemark{f} \nl 
\enddata
\tablenotetext{a}{LMC Cepheid $<$V$>$ from Udalski et al. (1999); \cite{Freed01} metallicity correction.}
\tablenotetext{b}{Caretta et al. (2000) LMC $<$V$_{RR}(0)>$  =  19.14 (includes +0.03, $<[$Fe/H$]>$ correction).}
\tablenotetext{c}{Weighted average. Error is standard deviation of the mean from four independent techniques.} 
\tablenotetext{d}{LMC Cepheid $<$V$>$ from Udalski et al. (1999); no metallicity correction.}  
\tablenotetext{e}{Weighted average. Error is standard deviation of the mean from five independent techniques.}               
\tablenotetext{f}{Udalski et al. (1999) LMC $<$V$_{RR}(0)>$  =  18.94 (+0.05, $<[$FeH$]>$
correction)  =  18.99} 
\end{deluxetable}
\end{center}

%
%

\clearpage

\clearpage

\clearpage
\clearpage

\clearpage

\clearpage

\end{document}